# A Refined Experience Sampling Method to Capture Mobile User Experience

**Mauro Cherubini**

Telefonica Research
Via Agusta, 177
08021 Barcelona, Spain
mauro@tid.es

**Nuria Oliver**

Telefonica Research
Via Agusta, 177
08021 Barcelona, Spain
nuriao@tid.es

**Abstract**

This paper reviews research methods used to understand the user experience of mobile technology. The paper presents an improvement of the Experience Sampling Method and case studies supporting its design. The paper concludes with an agenda of future work for improving research in this field.

## Introduction

Research on mobile technology has evolved rapidly during the last years stimulated by the market growth in this sector. However, research in this field has not grown harmonically, and the methodologies used were often not designed for the challenges posed by mobility. In most of the cases, researchers conducting experiments with mobile devices adapted methods, which were not mobile-specific, from other fields.

This paper reviews most commonly used research methodologies for understanding the mobile user experience (UX). As a result, we identify an opportunity for innovation. The main contribution of this paper is to propose a refinement of the Experience Sampling Method [12], whose virtues are discussed in detail in the review. In addition, this paper presents two case studies from our past work for which we discuss methodological challenges and lessons learned. A third case study illustrates the application of the method proposed in this paper. Informed by the work of Kjeldskov and Graham [32], the paper concludes with a discussion of challenges that the Mobile UX community might want to tackle in the near future.

**On mobile research methods and UX**

Järvinen described the approaches for empirical studies as either *creating* or *testing theories* [29, p.10]. *Theory-testing studies* are methods such as laboratory experiments, surveys, case-studies [46], field studies or tests. Some longitudinal studies belong also to this category [40]. Their underlying theory, model or framework is selected from the literature, or either developed or refined. On the other hand, *theory-creating approaches* include ethnographic methods [42], discourse analysis [1], and phenomenological studies [34]. The theory-testing studies belong more to the user experience research, which is directed at understanding "all aspects of the user's interaction with a product: how it is perceived, learned, and used" [36]. Therefore in the following review, we do not consider research frameworks like action research [16], grounded theory [41], basic research [45], etc. The discussion of the differences between these different approaches to research is outside the scope of this paper.

Depending on the situation under study, every research method threatens either the *internal* or *external validity* of a study. An experimental design is said to lack *internal validity* if results can be explained by factors that are additional to those explicitly incorporated in the design. Conversely, a study might lack *external validity* if the obtained results cannot be generalized to groups other than those that participated in the original study [29, p.54]. Another important dimension that helps differentiating the research methods is whether they require the subjects' active cognition, or interpretation, in the production of the results. If this is the case, the results are said to be *subjective* otherwise they are said *objective*. Objective data are usually preferred for experiments. On the contrary, the subjective measurement techniques are often the only direct means for the assessment of user opinions and preferences [14].

What follows is a review of empirical methodologies that can provide data to be analyzed through qualitative or quantitative techniques and that can shed light on human behavior, preferences, or advantages of use of a certain technology.

REVIEW OF QUALITATIVE MOBILE RESEARCH METHODS
<u>Survey research</u> involves aggregating information for answering a research question from a sample of the population using standardized instruments or protocols [33]. It includes administering a questionnaire, conducting field interviews, and using diary studies. One of the advantages of this technique is that it allows gathering large amounts of data with relatively little effort, supporting generalization of the results. However, survey research typically lacks internal validity. This technique has been used successfully in several studies involving mobile technology (see for instance the work of Frohlich and Murphy [20], or the technique combining interviews and surveys employed by Ichikawa et al. [27]). <u>Diary studies</u> of mobile interaction typically concentrate on macro events occurring during the interaction [24, 25]. However, the method poses a consistent demand on the participant who must interrupt the main activity to fill up the diary [38]. Additionally, the media used to record the events was found to have an impact on the subject's recognition and elicitation [47].

Another qualitative research method that is widely used for studying mobile technology are <u>observational</u>

*studies*, where the researcher follows the subject with a video camera, recording relevant aspects of the interaction context (sometimes this approach is called "shadowing" [10]). A methodology that belongs to usability testing that can be listed under this category is the "thinking aloud protocol" [17]. The advantages of observational studies are multiple: first, they do not require any additional effort from the participant in order to create the recordings and therefore they produce more objective data; second, the recordings can be coded and produce quantitative data [31]; third, these techniques can generate a large amount of grounded data in a relative short period of time. Major shortcomings of these methods are that there is no guarantee that the collected data is going to be representative, and that it is often difficult, intrusive, and improper to move around with devices recording the subject and relevant contextual elements [3]. To have the subject feel more comfortable in his/her activities a friend is asked to act as the researcher. This variation is called *pair observations* [28]. However, the friend might be improperly trained and therefore bias the data capture.

While the methods presented above strive to follow standardized and formal procedures to capture and analyze data, *cultural probes* aim at the opposite, namely taking into account the subjective interpretation of the research material by the participant and the researcher of the study [22]. They consist in a set of instructions and a recording device (e.g., diary, Dictaphone, disposable camera, etc) that is given to a number of participants, who interpret the instructions, create self-reports, and return the material to the researchers. Their main advantage is that they enable reaching environments that are difficult to behold directly, and they allow capturing aspects of human cognition that are difficult to observe otherwise (*e.g.*, feelings). In addition, cultural probes are less intrusive than observational studies and can be used in situations where other techniques might be disturbing [13, p.673]. In the vast environments of the city, variants of this technique were refined by Paulos and Jenkins, and called *urban probes* [48, p. 343]: "a fast-fail approach for asking early questions of urban computing in order to focus and influence future urban research and application choices".

Finally, *longitudinal studies* refer to techniques of collecting data, either qualitative or quantitative, at repeated intervals in time and comparisons are drawn on the different periods [40]. While the main advantage of this technique is that it reveals the dynamic nature of a variable, its main limitation is its inability to resolve issues of casual order [29]. Another important challenge of this technique is to control that no intervening factor might have biased the data between collection periods.

REVIEW OF QUANTITATIVE MOBILE RESEARCH METHODS
One of the main criticisms to the methods providing qualitative data are their potential biases derived by the subject's involvement in the capture of the observations [39]. Conversely, quantitative methods are based on automatic collection techniques. *Automatic logging*, for instance, is a technique that consists in recording events generated or sensed through mobile devices without the user's notice [9]. It offers a reliable way to gather consistent objective interaction data. However, this method is limited in its ability to capture the user's context and cognitive states.

Similarly, *controlled experiments* are particular setups where variables that are considered to have an impact on the research questions under study are reliably logged or manipulated. "Cognitive walkthroughs" are usability inspection methods that belong to this category [21]. In their conception, controlled experiments are usually conducted in controlled environments such as offices, hallways, laboratories, or simulators (*e.g.*, Bohnenberger *et al.* [4]), where logging and control of disturbing factors is easier. While the major advantage of this method is the high degree of control that the experimenters have on the variables under study, lab studies are usually criticized for their limited relation to the real world and unknown level of generalizability of the results outside the laboratory [45]. Mobility is very difficult to emulate in a laboratory setting.

The *quasi-experimentation* technique has been proposed [39] in an attempt to bring the design of controlled experiments in the actual environments where subjects actually use mobile technology. It employs a number of wearable micro-cameras in the effort of extending the range of recorded contextual information. It also structures the task that the subject is asked to complete so to have a finer degree of control over spurious factors [37]. While the technique achieves better internal validity, the necessary recording equipment and the imposed constrains on the behavior of the subject reduce its external validity.

Under particular circumstances, a single case can be used for theory-testing purposes. A *case study* is "an empirical enquiry that investigates a contemporary phenomenon within its real-life context, especially when the boundaries between phenomenon and context are not clearly evident" [46]. The data collected through case studies is grounded in natural settings [8]. Also, it is typically very rich and sometimes collected without a structured framework, resulting in incomplete or inconsistent datasets. According to Kjeldskov and Graham [32], case studies are particularly well suited for research focusing on describing and explaining a specific phenomenon, developing a hypothesis, or a theory.

All the techniques described in this section follow the principle of *logical deduction*, namely from theories to hypotheses, observations and generalizations. However, research on mobile technology is progressively adopting the principle of *logical induction*, namely from observations to generalization and theories [43]. The pervasiveness of ubiquitous technologies that are able to sense human behavior has led to huge amounts of raw data. *Datamining* techniques are applied to these datasets to learn recurring patterns [23] and to predict human behavior or future status of systems used by humans [15, 19]. The main limitation of this technique is its lack of internal validity, which is compensated by the large external validity of the so obtained results, given by the size of collected data. Following the same principle of logical induction, *simulations* are processes of obtaining the essential qualities of (real or planned) reality without reality itself [29]. Simulations enable dealing with realities that might exist within the range of possible realities depending on the values assigned or taken by configuration parameters [18, 44]. Their main shortcoming might consist of a poor model describing reality.

The names of the research methods represented in this graph are positioned according to their ability to produce results that fulfill the opposed requirements marked on the axes of the graph. This technique is usually employed in foresight research and it is used to individuate hotspots and "whitespots" (white areas). Two white areas are particularly evident in the first and third quadrant. The rESM proposed in this paper fills one of these empty spots.

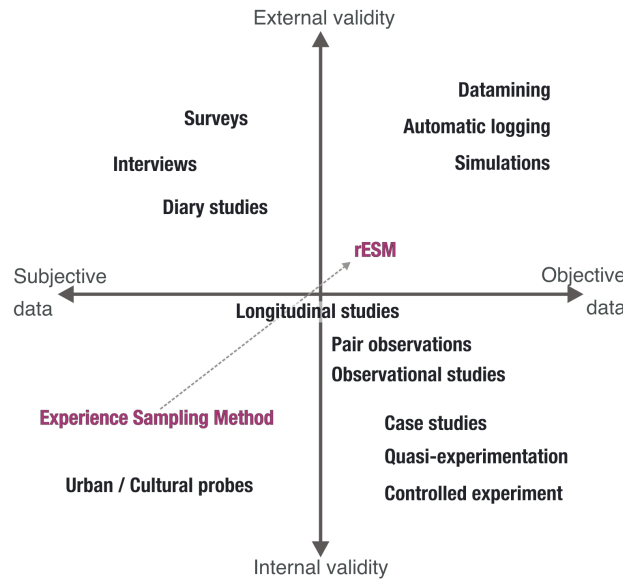

**figure 1.** Contrasting 2-dimension graph [11] of mobile research methods. This paper proposes a refined Experience Sampling Method (rESM) marked in purple in the graph.

**Figure 1** represents the methods discussed in this section in a contrasting 2-dimension graph where the techniques are mapped in the Cartesian space according to their ability to produce internally *vs.* externally valid results and collect subjective *vs.* objective data.

TOWARDS A REFINED EXPERIENCE SAMPLING METHOD
Qualitative research methods such as surveys, interviews, or diary studies rely on memory and are not suitable to study *micro-level interactions* (*e.g.*, why the user clicked on a certain button). Similarly, studies designed following quantitative research methods such as controlled experiments, quasi-experimentation, or observational studies have a profound impact on the natural context in which the interaction naturally happens. Therefore, it is difficult to use them for answering research questions targeting behavior on a micro-level of interaction. The *Experience Sampling Method* (ESM) was developed in the psychological field to tackle this issue [26]. It basically consists of randomly or semi-randomly sampling the user experience, usually by sending a message to the participant and asking him/her to answer a short questionnaire on a mobile device right at the moment when a relevant event is produced [12].

The main advantage of ESM is its ability to preserve the *ecological validity* of the measurements, defined by Hormuth [26] as: "the occurrence and distribution of stimulus variables in the natural or customary habitat of an individual". This method compares with recall-based self-reporting techniques –although recall delay is kept minimal– by "beeping" the subject in close temporal proximity to when a relevant event was produced. However and due to the level of involvement of each participant in the collection process, the method produces self-reported data. Particularly, Consolvo and Walker [12] describe how the alerts might be triggered using three modalities: randomly, on specified schedule, and triggered by the participants when a self-reported event of interest occurs. In the last two cases, the authors cautions on how the method might introduce a cognitive bias to the captured data.

Our proposal for a *refined ESM* (rESM) consists of: **1) the data collection process becomes automatic or semi-automatic through the information that can** **be automatically captured by mobile devices** (*e.g.*, the camera of the phone takes automatically a picture); and **2) this data collection is triggered by objective user-generated events** –non self-reported– (*e.g.*, the user makes a phone call). To illustrate this process imagine that we are interested in understanding the relationship between the linguistic content of an SMS and the geographical location of the recipient of the SMS. The rESM method can assist this research by logging the position of the recipient whenever the sender decides to write an SMS. rESM differs from automatic logging in two points: 1) the degree of involvement of the user in the collection process (the user might be aware and participate in the sampling); and 2) the nature of the contextual information to be gathered, which might be difficult to capture with automatic logging (*e.g.*, asking the user to take a picture of a furniture object nearby).

ESM was successfully employed in the Personal Server project [50]. The Personal Server is a mobile device that allows the user to store and access data and applications through interfaces found in the environment. For its design, the research team wanted to gain a better understanding of people's information needs while on the move and developed iESP, an open source package for running ESM on PalmOS. Through iESP, researchers sent 10 randomly scheduled alerts per day in a 12-hour time window (the study lasted 7 days). The deployed questionnaire included questions like: "Where are you?"; "Which of the following are available right now? –Printer, –Desktop Computer, -Laptop, –PDA", etc. Participants completed an average of 56 questionnaires. During the post-interviews, participants declared that primary reasons for not answering the questionnaires included being busy when they were alerted of the incoming questionnaire and not noticing the alert. Using rESM in this experiment, researcher could have achieved better results in three areas: 1) more questionnaires could have been completed; 2) filling the questionnaires would have required a smaller intervention from the subject as much information could have been sensed automatically (*e.g.*, the position of the user, or the presence of other Bluetooth-enabled equipment nearby); and 3) the probing requests could have been triggered in more relevant situations (*e.g.*, when the user changed position). iESP was later on revamped in the MyExperience tool [51], which allow subjective –as well as objective– measures of user's activities.

In summary, while ESM relies on the choice of the experimenter or the subject on when to deploy the probing requests, the rESM relies on objective system-generated events. This is the main contribution of this proposal.

**Case studies**
In our previous research on mobile technology we employed two techniques that provided us with satisfactory and unsatisfactory results respectively. In this section, we describe first two experiments (A and B) from our prior work focusing on what methodologies worked and what did not. Then we will describe a research (C) that we are about to conduct, for which the rESM might yield interesting results.

A- STRIVING FOR THE ECOLOGICAL VALIDITY
During summer 2006, the first author conducted a field study aimed at understanding the usefulness of sharing location-based annotations within a group of friends. The underlying hypothesis was that this technology could be appropriated by peers for sharing informal knowledge of the city they lived in. Twenty-one participants were given an application running on a mobile phone, named **STAMPS** [5, 7], with which they could create and share annotations (see **Figure 2** and **3**). The goal of the experiment was to understand: a) how an additional channel of communication could be integrated in existing communication strategies; b) what kind of messages participants would produce; and c) the relationship of the messages with the geographical locations to which they were attached. Therefore, it was decided to not give any incentives to the production of annotations and to not ask the participants to follow a particular procedure or script. Participants were left free to use the system the way they wanted, while the application logged automatically their interactions. The results were unsatisfactory [5, 7]. During four months participants produced an unequal number of annotations for a total of only 162 messages. A third of the participants did not produce any messages and logged in only one time (median). After an initial burst of interest during the initial two to three weeks, usage of the application decreased with few sporadic exceptions. As a consequence, the evaluation of the results had to be conducted on a small dataset, producing results that lacked external validity.

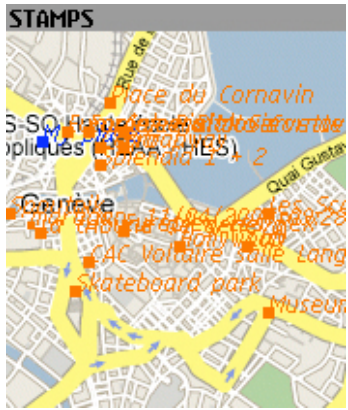

**figure 2.** STAMPS home screen. Annotations' anchor points are represented as little squares, and depending on the zoom level, some text of the annotation becomes also visible on the map.

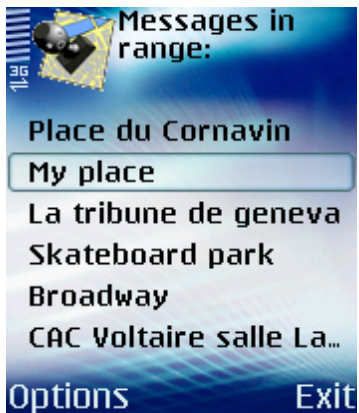

**figure 3.** List view of the annotations available in a certain region of the map.

In our experience, it is not productive to employ methodologies that aim at preserving the external validity of the results to conduct research on technology that is not part of the users' ecology –as in the case of STAMPS– mostly because it would not be possible to extend the findings to different groups. Our advice to researchers conducting studies on mobile technology not yet in common use is to frame the users' interactions in such a way that they would allow obtaining a consistent amount of evidences that support the subsequent analysis, and to spend effort in controlling external factors that could bias the results.

B- COMBINING FIELD STUDIES AND CONTROLLED EXPERIMENTS
More recently, we have been involved with a research project aiming at comparing textual and voice tags in supporting the indexing and retrieval of digital pictures captured through camera phones [6]. Our research question was whether one of these two modalities, or their combination, could be better suited for supporting the annotation of pictures while on the move, and to understand what modality could better support the subsequent retrieval of the pictures. Note that we needed to collect a multi-modal user-generated set of pictures and annotations for this research, as the application we employed was designed to be a personal system: only the user that generated the voice tags was meant to use them for retrieval. Also, answering our research question required studying the picture-taking activity that normally happens at sporadic intervals across different times and places. Therefore, we decided to use a combination of a field study and a controlled experiment. During the field study, twenty participants received a mobile phone with a prototype application, named **MAMI** [49, 53], which allowed them to collect pictures and associate multimodal tags to each image at the time of capture (see **Figure 4** and **5**).

We divided the participants in three experimental groups: seven participants could tag only with text, seven only with voice, and a final group of six could tag with both modalities. Participants could use the application during their holidays and the collected pictures and logs were used in the subsequent phase, where each of them was asked to retrieve a random sample of their collected pictures.

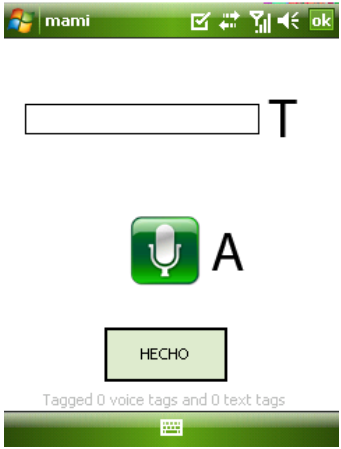

**figure 4.** MAMI indexing interface. Depending on the experimental condition elements marked A or T were not visible.

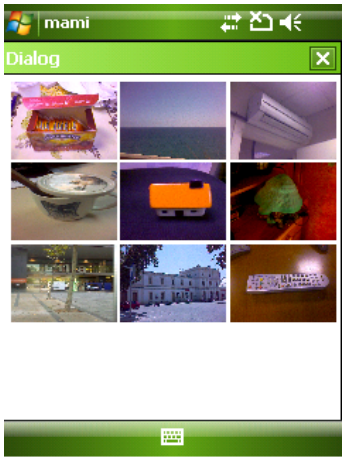

**figure 5.** MAMI query results interface.

This design was beneficial for many reasons: first, the division of the participants in different experimental groups already during the collection phase allowed us to compare the ability of the input modalities to support the production of the tags. Second, participants were familiar with the material used during the retrieval phase as they authored both the pictures and the tag descriptors used to retrieve them. Therefore, the second phase was an authentic personal retrieval task. Third, each methodology contributed with its best feature to support this research: the picture taking activity requires the naturalness of the real context and an extended period of time for capturing images. Therefore, it was studied through automatic logging. Conversely, measuring retrieval performance requires calculating time intervals which happen at a fast pace and that can be optimally controlled in the artificial setting provided by a controlled experiment. Fourth, interpretation of the results of the controlled experiment could be corroborated by data collected during the field trial, thus disambiguating operational hypotheses. In conclusion, we learned that each method has clear advantages in supporting specific research questions and that great synergies can be expected when combining different techniques in a single experimental design.

C- PROBING MENTAL MODELS

Many location-based messaging services that have been designed during the last years have failed at recognizing the dynamic process by which people construct and maintain the knowledge of the whereabouts of their peers. For example, Jung et al. [30] designed an application for sending SMS messages delivered under specific spatio-temporal constraints. They noticed how the system was used only when the users could predict in advance the context of delivery basing the inference on their knowledge of the receiver's schedule and movements in the city. The studies of Bentley and Meltcaf [2] reported relevant and related findings: peers have a good understanding of each other's whereabouts, but they are unsure of the exact transitioning times between locations. We consider this to be an experimental hypothesis that we want to verify. To this end, it is necessary to collect quantitative evidences during the subjects' daily life. Automatic logging would not be effective, because there is no way to automatically infer the subjects' mental models from the range of information that can be sensed through a mobile phone. A better choice would be to use ESM, by asking each participant (probe-peer) at regular intervals his/her best estimate for a friend's current position, and then ask this friend (test-peer) to disclose his/her position, thus providing the ground truth. However, data derived from ESM will suffer from a possible recall bias of both the probe-peer and the test-peer. Additionally, the provided answers, particularly those of the test-peer, might suffer also a bias of precision: the current location might be roughly communicated, delayed, or intentionally deceived. In these situations the rESM proposed in this paper might provide the best results, by combining automatic logging with probing questions. The rESM method would make it possible to query the probe-peer only when the answer would be mostly relevant for providing evidences for the research question (*e.g.*, when the test-peer is transitioning between two locations). Logging automatically the location of the test-peer might have multiple benefits: first, it can reduce the involvement of the test-peer in the data collection; second, position might be automatically sensed with objective precision; third, the same mechanism might provide reliable ground truth when it makes sense to send a probe request to a subject.

**Discussion**

In their review of mobile HCI research methods, Kjeldskov and Graham [32] concluded that: "given the fact that only few studies are based on a methodological foundation, it seems assumed that methodology matters very little in mobile HCI research" (p. 326). We could not agree more with these authors by stressing that setting the right experimental design has profound impacts on the results obtained by research [35]. Mobility poses a complete new set of challenges to practitioners, and methodologies that were developed in the past and in different disciplines need to be carefully adapted to this new theoretical field.

To this end, the objective of this paper was twofold: first, we reviewed the methods that are most often used in mobile UX research in order to identify opportunities for developing new research methodologies (the rESM was proposed); second, we introduced three case studies that could help ground lessons we learned in our prior work and explain how rESM could help answering specific research questions. From this contribution we would like to draw four implications for further work in this field.

1- INTEGRATION OF MULTIPLE METHODS
Our review of related work identified few examples of studies where authors successfully integrated different methodologies in the same research (especially between the qualitative and the quantitative categories). Our experience suggests that the integration of methods should be studied further as it might generate new research methods. Particularly, integration might be mostly beneficial in cross-modality verification of the results: data collected with one method might be further refined and verified through another method. However, integration might also result in a complete different methodology, as in the case of rESM proposed in this paper.

2- MOBILITY IS ABOUT CONTEXTUAL ISSUES
Little research has focused on identifying the role of contextual issues in the adoption of mobile technology. On the contrary, many studies focused on functionalities as if: "we already know what to build and which specific problems to overcome" [32]. Field studies and methods derived from ethnography could assist in identifying users needs and their subsequent translation into design opportunities. Mobile context needs to be studied in its complexity of dynamism and situated engagement, as Tamminen et al., did [52].

3- LACK OF DATASETS
Datamining the large datasets generated by mobile technology is a promising methodology [23]. However, this methodology is challenged by the lack of datasets and difficulty of aggregating user-generated data. The dataset aggregated in the Reality Mining project is a notable exception [15]. The major problem in this area is how to make information available to research while preserving the users' privacy. Specific solutions for making anonymous mobile information should be developed and discussed at community level.

### 4- CLAIM FOR A REFINED ESM

The refined ESM proposed in this paper might yield interesting results by providing objective, and semi-objective, measures that are also ecologically valid, and therefore, the so-obtained-measures might be generalized to other experimental groups. The reader can argue that datamining or logging is not equivalent to objective, since it's just a translation of action into countable observations. However, the goal of this paper is not to enter the recurrent debate between reductionism and holism. Our approach to this matter is merely practical and bound of the available and currently used methodologies.

Of course, the rESM method requires research in its ability to cope with the complexity of different experimental designs and its ability to provide solid and verifiable empirical results.

### Acknowledgements

We would like to thank Pierre Dillenbourg, who supervised the project described in the first case study, and Xavier Anguera who developed MAMI, the application described in the second case study. Also, we would like to thank Karen Church and Nicolas Nova for commenting a draft version of this work.